\documentclass{pasj00}
\draft

\begin{document}
\SetRunningHead{K. Fujisawa et al.}{Bursting Activity of the 6.7~GHz Methanol Maser in G33.641$-$0.228}

\title{Observations of the Bursting Activity of the 6.7~GHz Methanol Maser in G33.641$-$0.228}

\author{%
  Kenta \textsc{Fujisawa}\altaffilmark{1,2},
  Nozomu \textsc{Aoki}\altaffilmark{1},
  Yoshito \textsc{Nagadomi}\altaffilmark{1},
  Saki \textsc{Kimura}\altaffilmark{1},
  Tadashi \textsc{Shimomura}\altaffilmark{1},
  Genta \textsc{Takase}\altaffilmark{1},
  Koichiro \textsc{Sugiyama}\altaffilmark{1,3},
  Kazuhito \textsc{Motogi}\altaffilmark{1},
  Kotaro \textsc{Niinuma}\altaffilmark{1},
  Tomoya \textsc{Hirota}\altaffilmark{4},
  and
  Yoshinori \textsc{Yonekura}\altaffilmark{3}}
\altaffiltext{1}{Department of Physics, Faculty of Science, Yamaguchi University, Yoshida 1677-1, Yamaguchi-city, Yamaguchi 753-8512}
\altaffiltext{2}{The Research Institute of Time Studies, Yamaguchi University, Yoshida 1677-1, Yamaguchi-city, Yamaguchi 753-8511}
\altaffiltext{3}{Center for Astronomy, Ibaraki University, 2-1-1 Bunkyo, Mito, Ibaraki 310-8512}
\altaffiltext{4}{Mizusawa VLBI Observatory, National Astronomical Observatory of Japan, Hoshigaoka-cho 2-12, Oshu, Iwate 023-0861}
\email{kenta@yamaguchi-u.ac.jp}

%

\KeyWords{stars: formation --- ISM: individual (G33.641$-$0.228) --- masers --- radio lines: ISM} 

\maketitle

\begin{abstract}
We have observed bursting variability of the 6.7~GHz methanol maser of G33.641$-$0.228.
Five bursts were detected in the observation period of 294 days from 2009 to 2012.
The typical burst is a large flux density rise in about one day followed by a slow fall.
A non-typical burst observed in 2010 showed a large and rapid flux density enhancement
from the stable state,
but the rise and fall of the flux density were temporally symmetric
and a fast fluctuation continued 12 days.
On average, the bursts occurred once every 59 days,
although bursting was not periodic.
Since the average power required for causing the burst of order of 10$^{21}$~J~s$^{-1}$
is far smaller than the luminosity of G33.641$-$0.228,
a very small fraction of the source's power would be sufficient to cause the burst occasionally.
The burst can be explained as a solar-flare like event in which
the energy is accumulated in the magnetic field
of the circumstellar disk, and  is released for a short time.
However, the mechanism of the energy release and the dust heating process are
still unknown.
\end{abstract}

\section{Introduction}
The 6.7~GHz methanol maser (Menten 1991) is observed around high-mass young stellar objects
(e.g., Caswell et al. 1995; Minier et al. 2003; Xu et al. 2008; Breen et al. 2013).
This maser, thought to trace the circumstellar gas disk or outflow of the high-mass
star forming regions (e.g., Minier et al. 2000; De Buizer 2003; Sugiyama et al. 2014),
is characteristically variable (e.g., Goedhart et al. 2003, 2004). Goedhart et al. (2004)
reported various variation patterns, including periodicity,
in long-term monitoring observation of these events.
Fujisawa et al. (2012) (hereafter denoted paper I) discovered a bursting activity
of the maser in G33.641$-$0.228\footnote{The name G33.64$-$0.21 was used in the paper I,
but the sources is now usually called as G33.641$-$0.228.} (IRAS 18509+0027).
G33.641$-$0.228 is a high-mass star forming region at a near kinematic distance
of 4~kpc, and its bolometric luminosity is $1.2 \times 10^{4} L_{\solar}$. The 6.7~GHz methanol maser
of this source was first detected by Szymczak et al. (2000). 
The systemic velocity of this source is $61.5$~km~s$^{-1}$ (Szymczak et al. 2007).
It is reported in paper I that
the flux density of one spectral component (component II, $V_\mathrm{LSR}=59.6$~km~s$^{-1}$)
increased seven times with
a timescale of one day or less in a burst event, and then decreased over five days.
Very long baseline interferometry (VLBI) observation revealed the size of the bursting region
as less than 70 AU. From these observations, the burst is considered as
an explosive phenomenon occurring within the circumstellar disk,
probably sourced from the magnetic energy therein.

Paper I reported two bursts of the 6.7~GHz methanol maser in G33.641$-$0.228 during a 108-day observation period in 2009.
Szymczak et al. (2010) detected another large flux variability in MJD = 55160 (November 25, 2009).
To understand the mechanism of the burst, its properties such as occurrence frequency and velocity drift should be measured
from long-term observations. As a follow-up to the 2009 observation,
we monitored G33.641$-$0.228 in 2010, 2011 and 2012.
In this paper, the observations and results are presented in section 2,
and discussed in section 3. The paper concludes with section 4.

\section{Observations and Results}
\subsection{Observations}

G33.641$-$0.228 was monitored during the summer or autumn of 2009, 2010, 2011, and 2012.
The 2009 observation was reported in paper I. In 2010, this source was monitored daily
from July 29 to September 23, except for two periods of no observation;
from August 9 to August 16 and from September 11 to September 20. In total,
38 observations were made during the 57-day period. The 2011 observations
were conducted two times every three days over an 83-day period covering September 12
to December 3, yielding 41 observations. The 2012 observation covered 46 days
from September 27 to November 11. Over this period, G33.641$-$0.228 was monitored
almost daily, and 43 observations were made. The observation periods and numbers
are summarized in Table \ref{table1}. The observation system, calibration method,
and accuracy of the data are reported in paper I.
The observing bandwidth, number of frequency channels, and velocity resolution
were 4~MHz, 4096, and 0.044~km s$^{-1}$, respectively. The $1\sigma$ rms noise level is 1.4~Jy
with an integration time of 14~minutes. In 2012, the observation bandwidth was
8~MHz and the number of frequency channels was 8192. Other parameters were unaltered
from the 2011 observation.

\begin{table*}[htbp]
\begin{center}
\caption{Observation dates of G33.641$-$0.228.}
\label{table1}
\begin{tabular}{ccccccc}
\hline
\hline
Year    & \multicolumn{2}{c}{Start date}& \multicolumn{2}{c}{End date} & Observing period & Number of observation \\
        & date       & MJD   & date     & MJD   & (days)               &                       \\
\hline 
2009    & Jul 2      & 55014 & Oct 17   & 55121 & 108                  & 62                    \\
2010    & Jul 29     & 55406 & Sep 23   & 55462 &  57                  & 38                    \\
2011    & Sep 12     & 55816 & Dec 3    & 55898 &  83                  & 41                    \\
2012    & Sep 27     & 56197 & Nov 11   & 56242 &  46                  & 43                    \\
\hline 
\end{tabular}
\end{center}
\end{table*}

\subsection{Results}
\begin{figure}
  \begin{center}
    \FigureFile(160mm,160mm){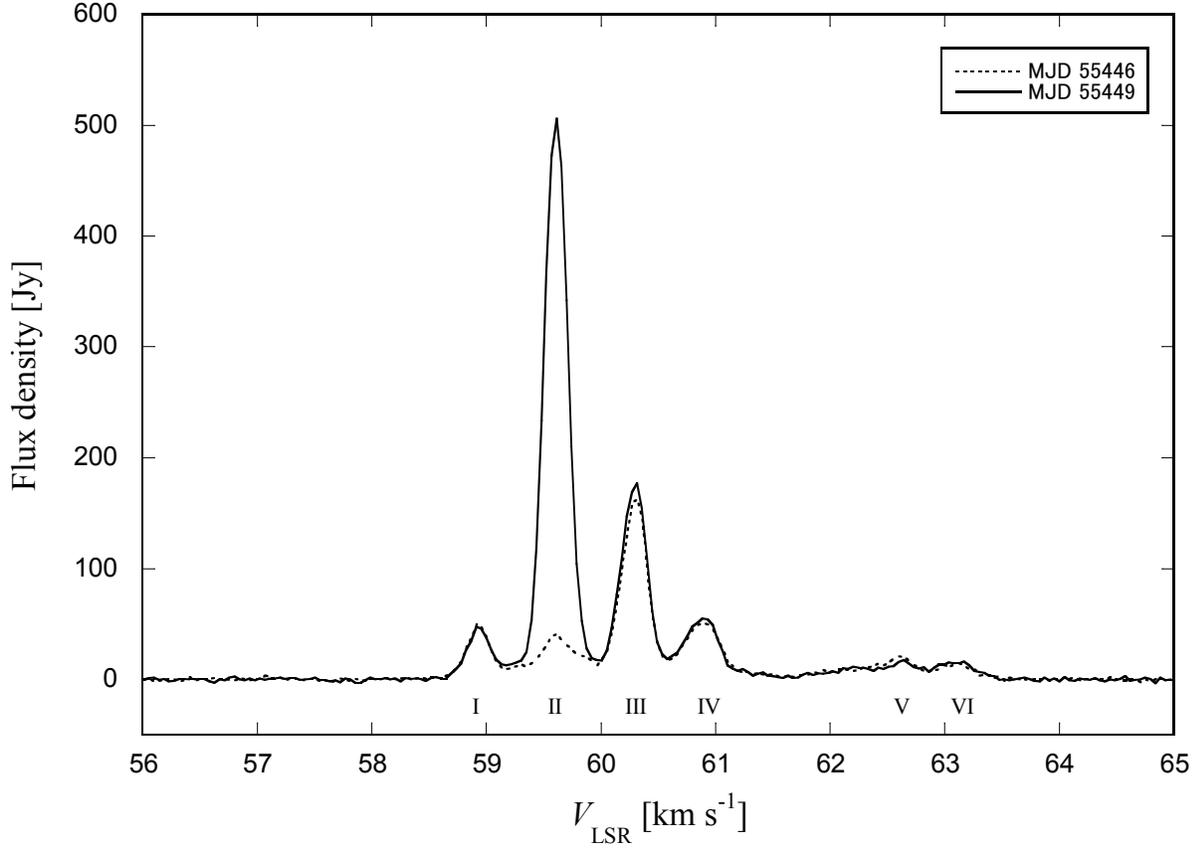}
  \end{center}
  \caption{Spectra of G33.641$-$0.228. 
The dotted and solid lines show the spectra at MJD = 55446 (September 7, 2010; just before a burst), and 55449 (September 10, 2010; at the time of the burst), respectively. The six spectral components are clearly seen.
Only the spectral component II at $V_\mathrm{LSR}=59.6$~km~s$^{-1}$ dramatically changed during the three days.
}\label{fig:fig1}
\end{figure}

As described in paper I, five sharp spectral peaks (components I$-$V)
and one broad peak (VI) are seen in the spectrum of G33.641$-$0.228.
Although the flux varied, the six spectral components have been clearly
identified over the four years from 2009 to 2012. Two spectra of a typical burst
event are shown in Figure \ref{fig:fig1}; September 7, 2010 (MJD = 55446; just before a burst,
dotted line) and September 10, 2010 (MJD = 55449; at the time of the burst, solid line).
Since the bursting occurred only in component II and no correlated variability
appeared in the other components throughout the observation period,
we mainly focus on the variability of component II throughout the remainder of this paper.

\begin{figure}
  \begin{center}
    \FigureFile(150mm,150mm){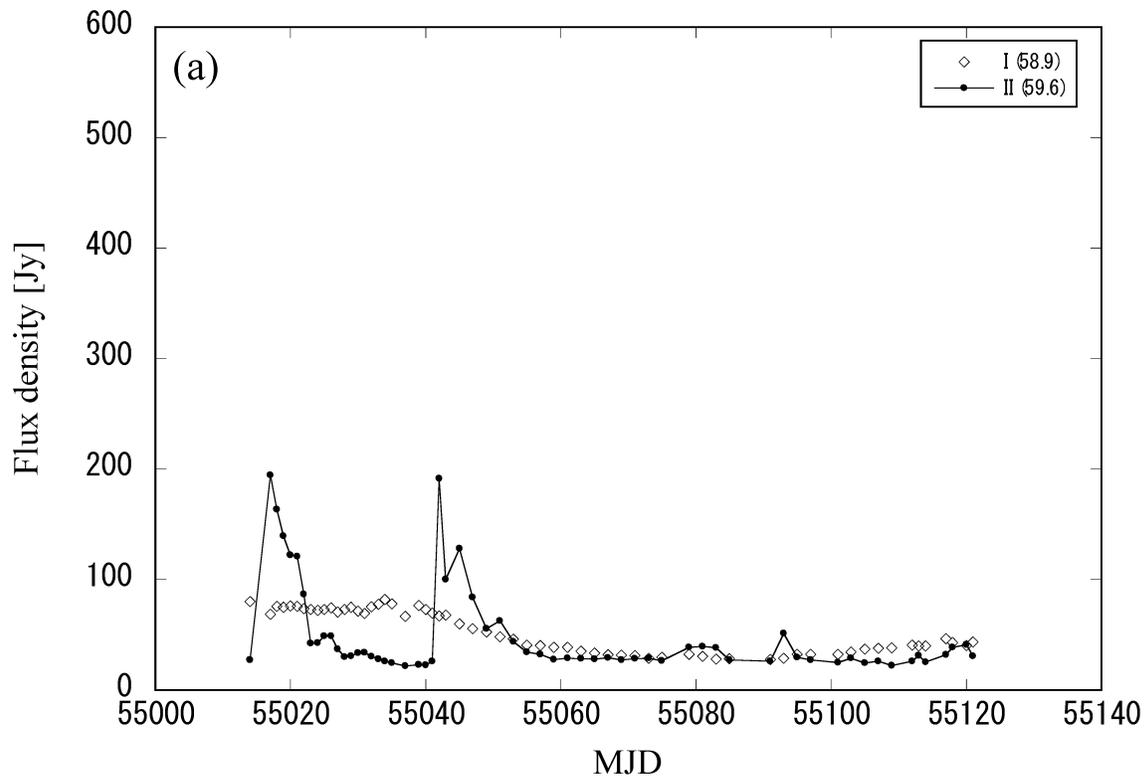}
    \FigureFile(150mm,150mm){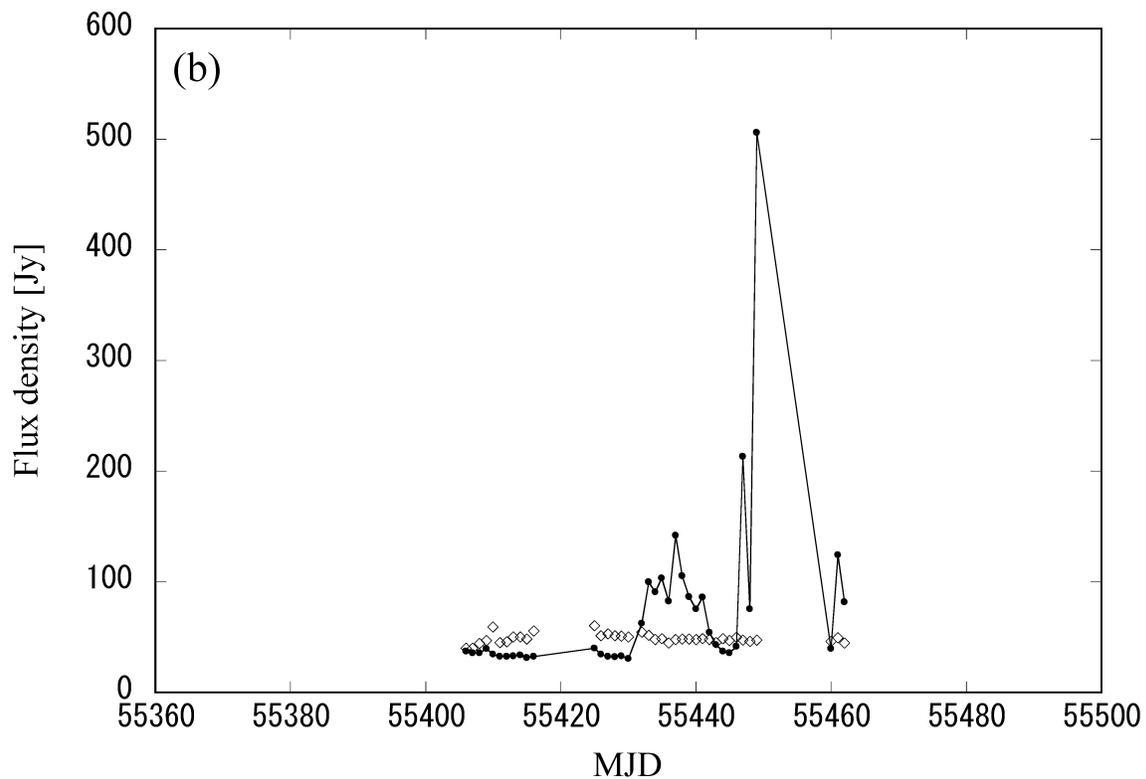}
  \end{center}
  \caption{Variations of the spectral components I and II observed in 2009, 2010, 2011 and 2012, respectively (a--d). Filled circles with solid lines indicate component II, and open diamonds indicate component I. The scales of the horizontal and vertical axes are set to be the same in each panel.
}\label{fig:fig2}
\end{figure}

\setcounter{figure}{1}

\begin{figure}
  \begin{center}
    \FigureFile(150mm,150mm){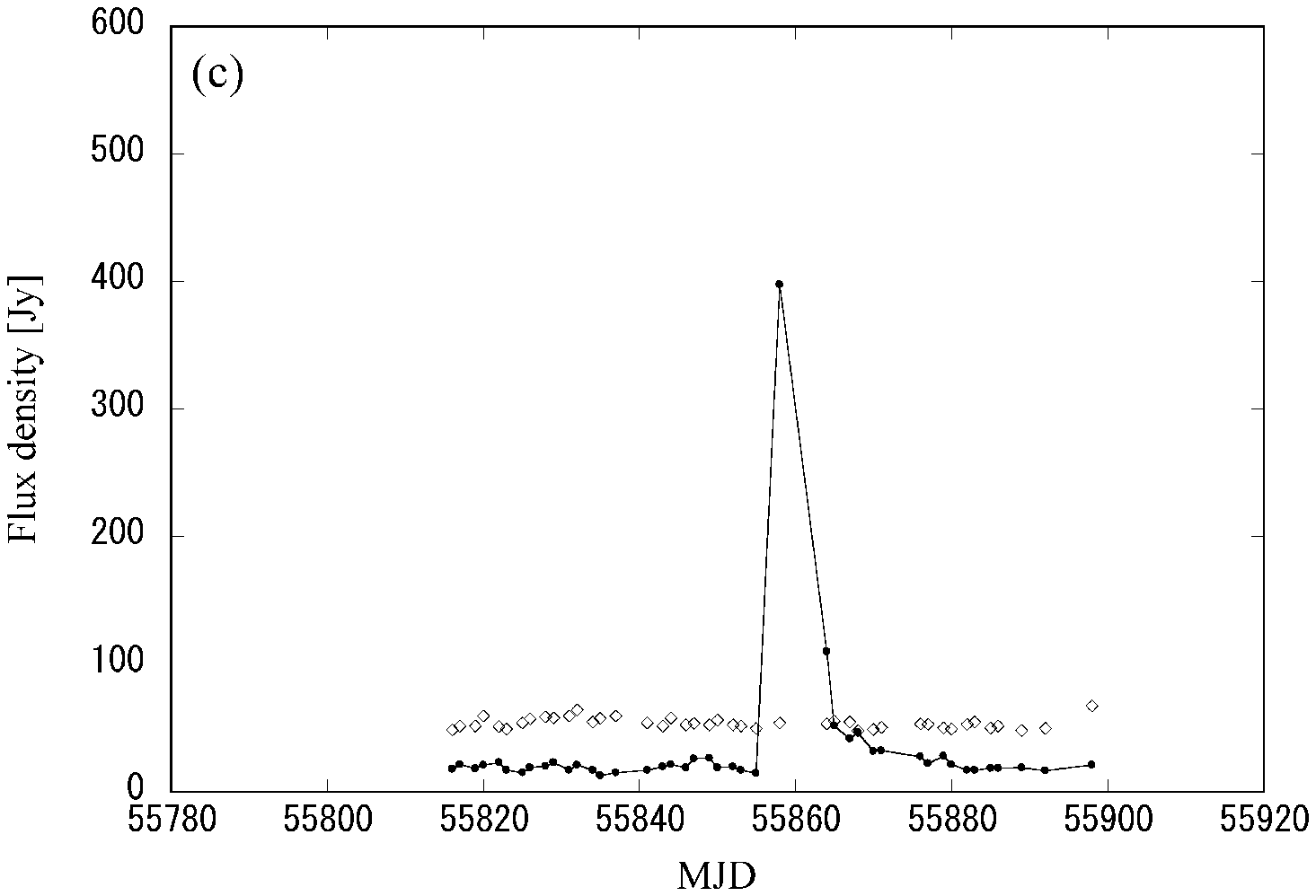}
    \FigureFile(150mm,150mm){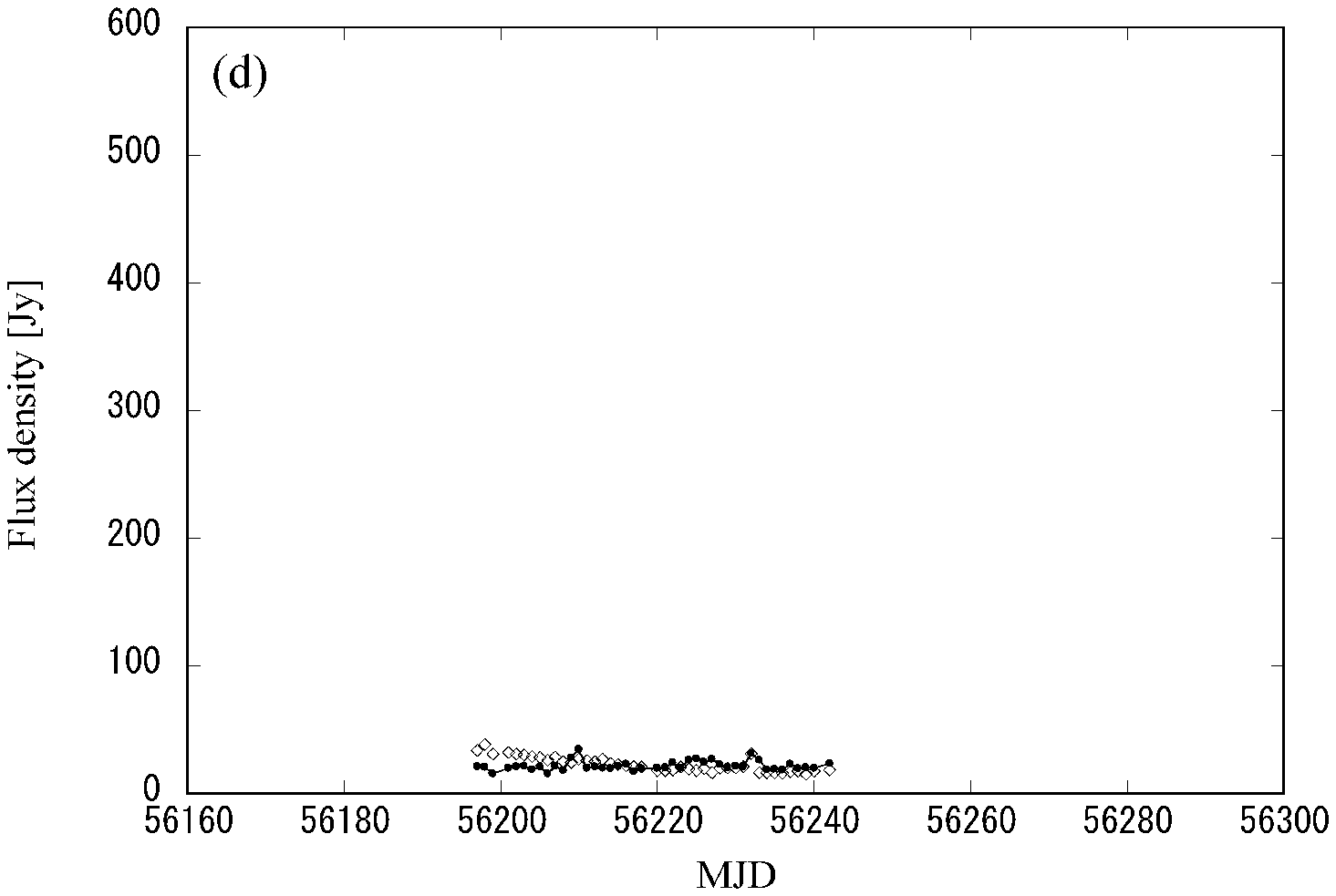}
  \end{center}
  \caption{Continued.
}
\end{figure}

\begin{figure}
  \begin{center}
    \FigureFile(160mm,160mm){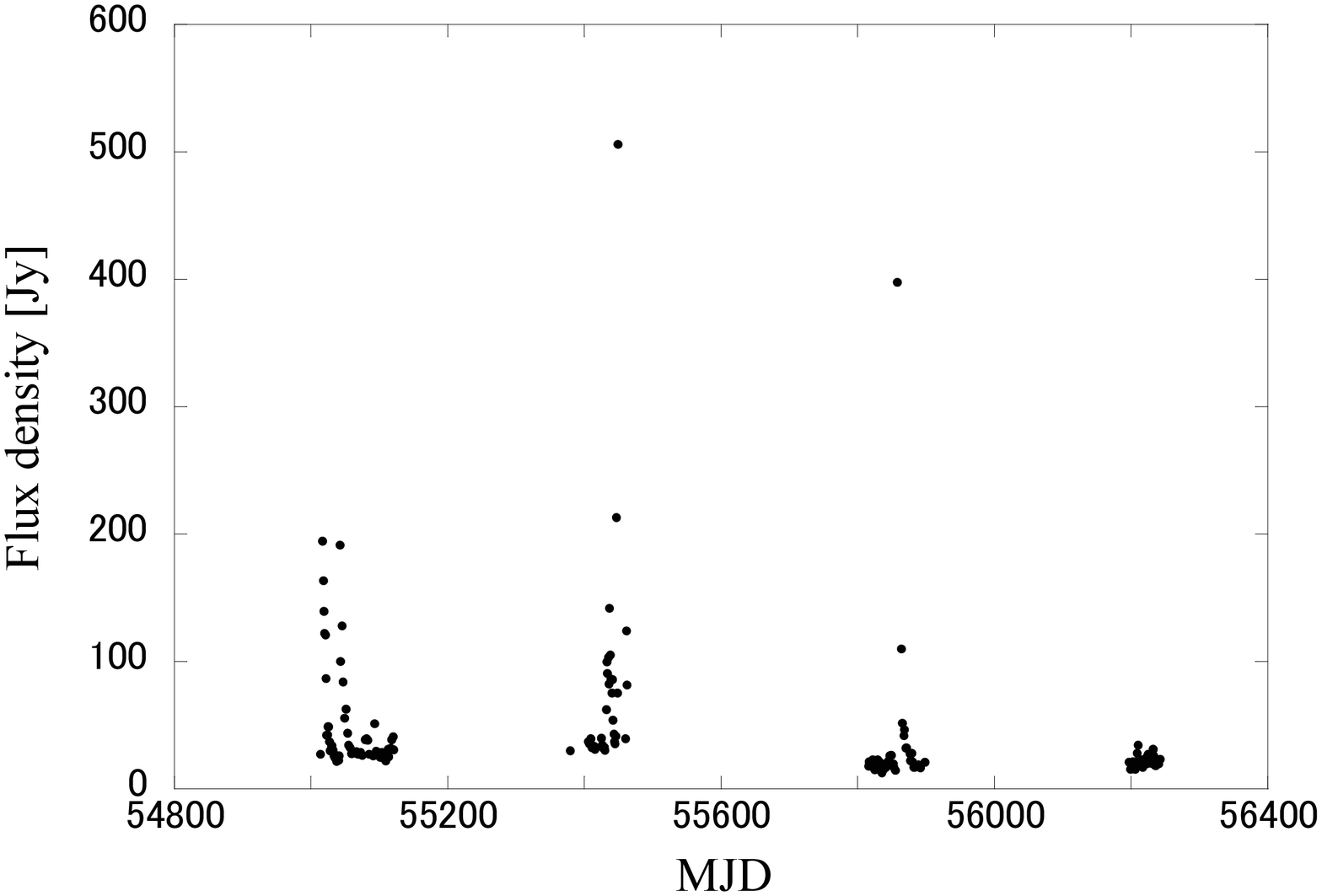}
  \end{center}
  \caption{Variations of the spectral components II observed from 2009 to 2012.
}\label{fig:fig3}
\end{figure}

The variations of the flux density of spectral component II and component I for comparison
observed in 2009, 2010, 2011 and 2012 are shown in Figure \ref{fig:fig2} a$-$d, respectively.
Typical bursts were detected
two times in 2009 (Figure \ref{fig:fig2}a, reported in paper I). The flux density of component II
($V_\mathrm{LSR} = 59.6$~km~s$^{-1}$) increased by seven times within one or three days, and decreased
over the following five days. 
In the last 40 days of the observation period,
the flux density of component II fluctuated with
a timescale of one or two days with an amplitude of 20~Jy.
For component I, relatively rapid fall with an e-folding time of 16~days was observed from
MJD = 55039. The variability like this,
however, has been seen in other source (e.g., Sugiyama et al. 2008).
The timescale of 16~days is much longer than that of the burst of component II.
In 2010 (Figure \ref{fig:fig2}b),
two bursts were detected. 
Light curve of the first burst is not a typical one.
The rise and fall of this burst were almost
temporally symmetric from MJD = 55432 to 55443. From its initial detection,
the flux density rose to its maximum after six days, and returned to the stable state
over the next six days. 
The variation was not smooth but a fast fluctuation with a time scale of one day.
At its peak, the flux density reached 142~Jy,
four times that of the stable state immediately before the burst.
After the first burst,
the flux density returned to
41~Jy on MJD = 55446. The following day (MJD = 55447), a large burst of flux
density 213~Jy occurred, then falling to 75~Jy one day later (MJD = 55448).
Next day (MJD = 55449), the flux density was again rapidly increased, reaching
to 506~Jy. This rapid increase in flux density
over a short period (one day) is a typical
characteristic of the burst of this source.
Unfortunately, the observation was halted for ten days
after the burst, and the behavior of the variation could not be observed.
The flux density decreased to 39~Jy in the observation 
after the burst (MJD = 55460), then increased to 124~Jy in the following day.
Whether this enhancement is related to the earlier large bursts is unclear.
One typical burst was observed in 2011 (Figure \ref{fig:fig2}c). The flux density of 15~Jy
on MJD = 55855 was increased to 398~Jy two days later (MJD = 55857),
then decreased over the next five days. The timescales of the rise and fall,
two days and five days respectively, were similar to those of
the two bursts observed in 2009.
Note that, because the flux density varies on short timescales,
probably less than one day,
the variation profile of each burst is not known in detail.
Even the decreasing occurred over one day in 2010 observation.
The details of the variability require sub-daily observations,
and we planned a sub-daily observation in 2012 (Fujisawa et al. 2014).
Unfortunately, no burst was observed in 2012 observation (Figure \ref{fig:fig2}d),
but small flux enhancements of about 10~Jy with one-day timescale were observed at MJD = 56210 and 56232.

\section{Discussions}

The typical light curve of a burst is characterized by a rapid increase in flux density
of more than 100~Jy over one day, and the following decrease over the next five days.
At the non-typical burst in 2010, the flux density fluctuated with the timescale of one day,
and the fluctuation lasted 12 days.
Although the light curve was different from that of the typical burst,
its properties are similar to those of the typical one:
the burst occurred only in component II;
the timescale of the variation is as short as one day;
the flux enhancement occurred from and returned to the stable state;
the e-folding time of the decreasing phase of the flare was 5.7 days, which is
close to that of the typical burst.
This non-typical burst probably is constituted of successive small bursts.
The flux fluctuation with an amplitude of 10$-$20~Jy were observed in 2009
and 2012 observations. These fluctuations also might be interpreted as small bursts.

As well as the result in 2009, only component II showed fast variability over four years of observation.
The bursting activity was observed at least for three years. 
The flux variation of component II from 2009 to 2012 are collectively shown in Figure 3.
In 294 days of observation over four years, five bursts were observed;
the occurrence frequency is once every 59 days on average, but with no regularity.
Two bursts were occurred by 25 days interval in 2009, whereas the third burst did not
occur within 60 days after the second.
In paper I, it is suggested that the burst is
generated in the following processes:
the energies of accretion, rotation, or radiation of G33.641$-$0.228
are accumulated in the local magnetic field in the accretion disk,
and the energy is explosively released in a short time like solar flare.
The released energy is converted into the heat of
gas and dust near the bursting component,
and finally the maser emission around the release point is suddenly strengthened.
In solar flare, energy is slowly accumulated
in the magnetic field on the solar surface and released in a short time by means of reconnection
(Tsuneta 1996).
Two characteristics of the solar flare, i.e.,
the solar flare does not have periodicity, and the energy release of the solar flare
occurs in the local region on the solar surface, are applicable to the maser burst.
The similarity of the solar flare and the maser burst
supports the model of the burst proposed in paper I.
Assuming the energy required for one burst as $6 \times 10^{27}$~J (paper I),
and given the burst occurrence frequency, the average power causing the burst would be
approximately $10^{21}$~J~s$^{-1}$, which is much smaller than the luminosity of this source
($4.6 \times 10^{30}$~J~s$^{-1}$). Therefore, if a very small fraction of the source's
energy is accumulated into the magnetic field in the accretion disk and released occasionally,
the energy would be sufficient to produce the burst.

\begin{figure}
  \begin{center}
    \FigureFile(160mm,160mm){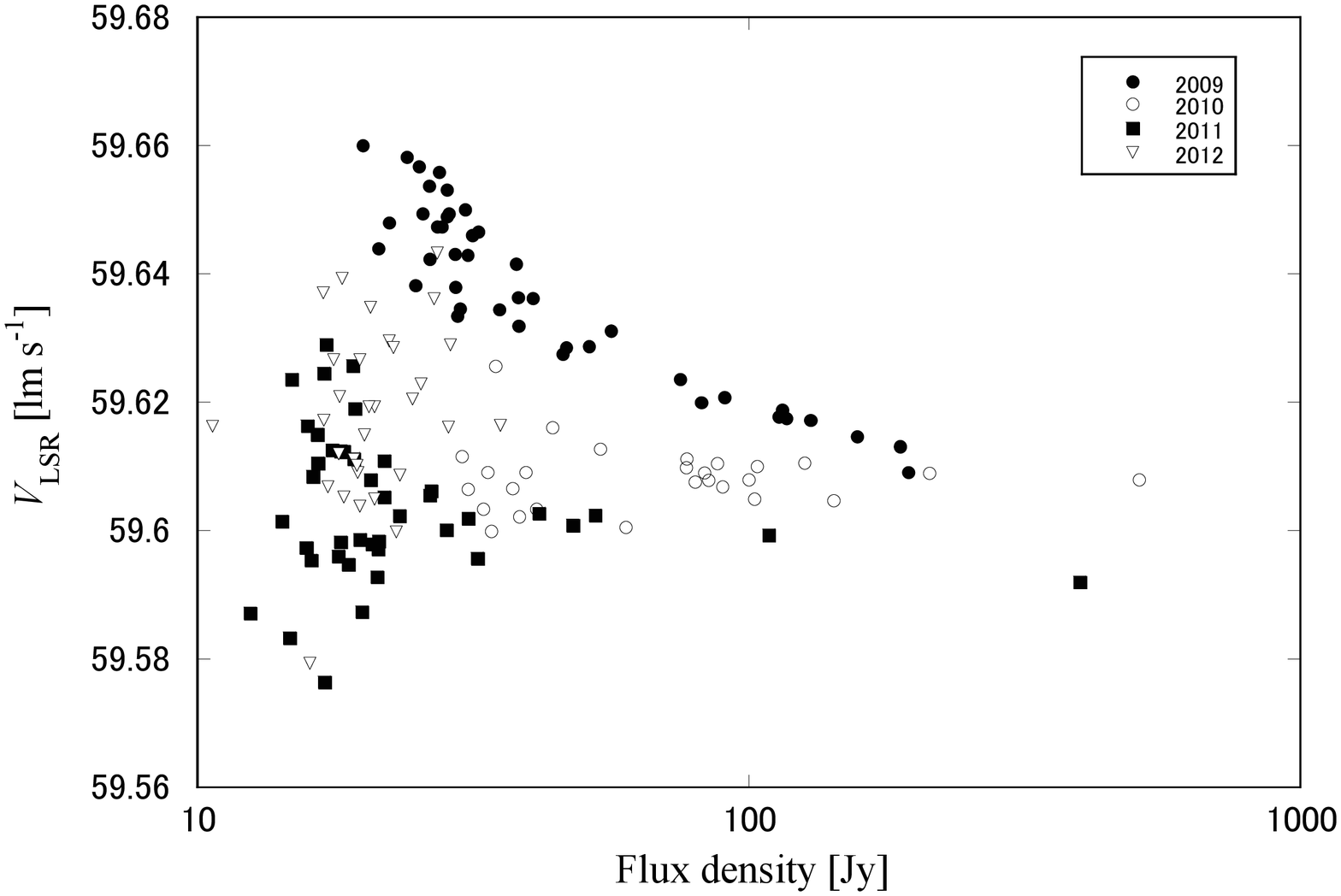}
  \end{center}
  \caption{Peak velocity with respect to flux density of component II. Peak velocity is negatively correlated with flux density in the 2009 data (filled circles), but no clear correlation appears in the data of 2010 (open circles) and 2011 (filled squares).
}\label{fig:fig4}
\end{figure}

In paper I, a decreasing peak velocity of component II was reported when the flux density of component II
increased in the 2009 bursts. 
As discussed in paper I, the velocity change in 2009 could be explained as
that the bursting component is a superposition of a stable component and a bursting
component with slightly different velocities.
In fact, Bartkiewicz et al. (2012) reported that there were two spectral component, {\it k} and {\it l},
at the velocity range of the bursting component from their VLBI observation carried out on 2003 June.
The velocity $V_\mathrm{LSR} = 59.6$~km~s$^{-1}$ of component {\it l} is closer to that of the bursting component
(59.6~km~s$^{-1}$)
and is slightly smaller than 59.8~km~s$^{-1}$ of {\it k}.
Since the observed peak velocity was decreased in the bursts in 2009,
the component {\it l} and {\it k}
would correspond to the bursting component and the stable component, respectively.
In some periods, the flux density of component II was quite stable.
Examples of stable periods are the middle of 2009 (MJD = 55059$-$55075),
the first half of 2010 (55406$-$55430).
The stable state would be explained
as the very low state of component {\it l} while the stable component {\it k} is visible.
In the 2010 and 2011 observations, 
the peak velocity at the non-bursting state was $\sim 59.60$~km~s$^{-1}$
which was smaller than that of observed in 2009,
and it did not change during the bursts (Figure \ref{fig:fig4}).
This could be explained as a velocity drift of component {\it k}, or
an appearance of another weak spectral component at 59.60~km~s$^{-1}$ 
and disappearance of component {\it k}.
In 2012, the observed velocities were around 59.62~km~s$^{-1}$.

Observations of various molecular lines
(water maser, HCO$^{+}$(1$-$0), H$^{13}$CO$^{+}$(1$-$0),
OH masers; Szymczak et al. 2007, Szymczak \& Gerard 2004)
in G33.641$-$0.228 is discussed by Bartkiewicz et al. (2012)
as an evidence of existance of an outflow and a shock in this source.
A part of water maser spot exists close to
the spot of methanol maser which causes the burst (Bartkiewicz et al. 2011, 2012). 
However, it is very unlikely that a shock cause the burst.
Although the velocity range of these lines coincides with that of the methanol maser,
spatial distribution of the methanol maser is elliptical and it probably traces
the disk rather than the outflow and/or shock region.
The peak velocity of the bursting component changed less than 20~m~s$^{-1}$ for three years,
which means that the bursting region has been kinematically stable even its excitation state changes rapidly.

On the other hand, the mechanisms of the releasing of local magnetic energy and heating of dust are unknown.
Also, why such bursts are unique to this source remains unclear. 
Since the observations reported here are obviously under sampling to the fast variability of the burst,
the detailed light-curve at the time of a burst is not yet known.
If the detailed light-curve at the time of a burst is observed,
some restrictions would be put on the energy release and excitation mechanisms.
Moreover, this source was detected at 12.2 GHz methanol maser (B\l aszkiewicz and Kus 2004).
If it is observed how the 12.2 GHz maser changes at the time of the 6.7 GHz burst,
an excitation state during the burst could be investigated in detail.

\section{Conclusions}
We have reported bursting variability in the 6.7~GHz methanol maser of G33.641$-$0.228.
Five bursts were detected throughout 294 days of observations performed from 2009 to 2012.
Among six spectral components, bursts occurred only in spectral component II;
the other components remained quiescent during bursting.
The typical timescales of the rise and fall phases were one and five days, respectively.
Bursting was not periodic, but occurred once every 59 days on average.
The burst mechanism can be explained as rapid release of magnetic energy in
the gas disk surrounding the high-mass young stellar object.
However, the mechanism of the burst and why the burst is unique to this source remains unknown.

\bigskip

The authors thank Professor Mashiyama for his encouragement of this study.
The authors also thank
National Astronomical Observatory of Japan, KDDI corporation for supporting
the Yamaguchi 32-m radio telescope.


\end{document}